\documentclass[12pt]{article}

\begin{document}

\title{\textbf{Generalized Intelligent States for an Arbitrary Quantum 
System}%
\vspace{0.5cm}}
\author{A. H. EL Kinani$^1$ and M. Daoud$^2$\vspace{0.5cm} \\
$^1$L.P.T, Physics Department, Faculty of Sciences, \\
University Mohammed V, P.O.Box 1014\\
Rabat, Morocco.\vspace{1cm}\\
$^2$L.P.M.C, Physics Department, Faculty of Sciences, \\
University Ibn Zohr, P.O.Box 28/S\\
Agadir, Morocco.}
\date{}
\maketitle

\begin{abstract}
Generalized Intelligent States (coherent and squeezed states) are derived
for an arbitrary quantum system by using the minimization of the so-called
Robertson-Schr\"odinger uncertainty relation. The Fock-Bargmann
representation is also considered. As a direct illustration of our
construction, the P\"oschl-Teller potentials of trigonometric type will be
shosen. We will show the advantage of the Fock-Bargmann representation in
obtaining the generalized intelligent states in an analytical way. Many
properties of these states are studied.
\end{abstract}

\newpage\

\section{\textbf{Introduction}}

The well known coherent states of the harmonic oscillator have turned out to
be one of the most useful tools of quantum theory $\left[ 1-4\right] $.
Introduced long ago by Schr\"odinger $\left[ 5\right] $, they were employed
by Glauber and other authors in quantum optics $\left[ 6-8\right] $. Further
developments of the subject made it possible to set up some specific
definitions, which are applicable to various physical systems. They were
discussed in connection with exactly solvable models and non-linear algebras
$\left[ 9-12\right] $ as well as deformed algebras $\left[ 13\right] .$

Recently, a construction of coherent states for an arbitrary quantum system
has been proposed by Gazeau and Klauder $\left[ 14\right] $ (see also $%
\left[ 15\right] $ and $\left[ 16\right] $). An interesting illustration of
this construction was given in $\left[ 17\right] $ for a particle trapped in
an infinite square well and in P\"oschl-Teller potentials of trigonometric
type $\left[ 18\right] $. The Gazeau-Klauder coherent states $\left[
14\right] $ are eigenvectors of the annihilation operator.

On the other hand, the squeezed states of electromagnetic field have
attracted due attention in the last decade (see for instance the references 
$%
\left[ 3,18\right] $). In the recent years, considerable interest has also
been devoted to the squeezed states for spin components $\left[ 19-20\right]
$, the number and phase operators $\left[ 21\right] $, the generators of the
algebras $su(2)$ and $su(1,1)$ $\left[ 22-24\right] $ and the supersymmetric
oscillator $\left[ 25\right] $.

The aim of the present work is to consider some general properties
of generalized intelligent states for an arbitrary quantum system.
These states minimize $\left[ 26\right] $ the
Robertson-Schr\"odinger uncertainty relation $\left[ 27-28\right]
$ and generalize the Gazeau-Klauder coherent states $\left[
14\right] $. coherence and squeezing is discussed throughout this
paper.

The paper is organized as follows. In section 2, we recall the main results
concerning the states minimizing the Robertson-Schr\"odinger uncertainty
relation and some useful formulae which are relevant in the study of
coherence and squeezing of these states. The generalized intelligent states
minimizing the Robertson-Schr\"odinger uncertainty relation are explicitly
computed in section 3 and we show that they generalize the Gazeau-Klauder
coherent states in some special cases, which will be discussed. In section
4, we introduce the Fock-Bargmann realization of the Gazeau-Klauder coherent
states by means of which we construct, in section 5, the P\"oschl-Teller
intelligent states. Coherence and squeezing of such states are also
considered. Conclusions and concluding remarks are given in section 6.

\section{$\mathbf{Robertson}-\mathbf{Schr\ddot odinger\ Uncertainty}$ $%
\mathbf{Relation}$}

Choose a Hamiltonian $H$ with a discrete spectrum which is bounded below,
and has been adjusted so that $H$ $\geq 0$. For convenience, we assume that
the eigenstates of $H$ are non degenerate. The eigenstates $\left| \psi
_n\right\rangle $ of $H$ are orthonormal vectors and they satisfy

\begin{equation}
H\left| \psi _n\right\rangle =e_n\left| \psi _n\right\rangle  \label{aq}
\end{equation}
In a general setting, We also assume that the energies $e_0,e_1,e_2,...$ are
positive and verify $e_{n+1}>e_n$. The ground state energy $e_0=0$.
Therefore, there is a dynamical algebra generated by lowering and raising
operators $a^{+}$(creation operator) and $a^{-}$(annihilation operator) such
that the Hamiltonian $H$ can be factorized as:

\begin{equation}
H=a^{+}a^{-}  \label{as}
\end{equation}
The action of the operators $a^{+}$ and $a^{-}$on the $\left| \psi
_n\right\rangle $ are given by

\begin{eqnarray}
a^{-}\left| \psi _n\right\rangle &=&\sqrt{e_n}e^{i\alpha
(e_n-e_{n-1})}\left| \psi _{n-1}\right\rangle  \nonumber \\
a^{+}\left| \psi _n\right\rangle &=&\sqrt{e_{n+1}}e^{-i\alpha
(e_{n+1}-e_n)}\left| \psi _{n+1}\right\rangle \hspace{0.7cm},\hspace{0.8cm}%
\alpha \in \mathbf{R}  \label{ad}
\end{eqnarray}
implemented by the action of $a^{-}$ on the ground state $\left| \psi
_0\right\rangle $

\begin{equation}
a^{-}\left| \psi _0\right\rangle =0  \label{af}
\end{equation}
The exponential factor appearing in all expressions produces only a phase
factor, and will be significant for the temporal stability of the
generalized intelligent states, which we will construct in the following. 
\\%
The commutator of $a^{+}$and $a^{-}$is defined by

\begin{equation}
\left[ a^{-},a^{+}\right] =G\left( N\right) \equiv G  \label{ag}
\end{equation}
where the operator $G\left( N\right) $ is defined by its action on states $%
\left| \psi _n\right\rangle $

\begin{equation}
G(N)\left| \psi _n\right\rangle =\left( e_{n+1}-e_n\right) \left| \psi
_n\right\rangle  \label{ah}
\end{equation}
It is diagonal with eigenvalues $\left( e_{n+1}-e_n\right) .$ We define the
operator number $N$ as
\begin{equation}
N\left| \psi _n\right\rangle =n\left| \psi _n\right\rangle  \label{aj}
\end{equation}
which is in general different from the product $a^{+}a^{-}$($=H$). We can
see that it satisfies the following commutation relations

\begin{equation}
\left[ a^{-},N\right] =a^{-}\hspace{1cm}\left[ a^{+},N\right] =-a^{+}
\label{ak}
\end{equation}
Using $a^{+}$ and $a^{-},$ we introduce two hermitian operators
\begin{equation}
W=\frac 1{\sqrt{2}}\left( a^{-}+a^{+}\right) \hspace{1.0in}P=\frac 
i{\sqrt{2}%
}\left( a^{+}-a^{-}\right)  \label{al}
\end{equation}
which satisfy the commutation relation
\begin{equation}
\left[ W,P\right] =iG  \label{am}
\end{equation}
The operator $G,$ in general, is not necessarily a multiple of the unit
operator. It is well known that for two hermitian operator $W$ and $P$
satisfying the noncanonical commutation relation $\left( 10\right) ,$ the
variances $\left( \Delta W\right) ^2$ and $\left( \Delta P\right) ^2$
satisfy the Robertson-Schr\"odinger uncertainty relation
\begin{equation}
\left( \Delta W\right) ^2\left( \Delta P\right) ^2\geq \frac 14\left(
\langle G\rangle ^2+\langle F\rangle ^2\right)  \label{zq}
\end{equation}
where the operator $F$ is defined by:
\begin{equation}
F=\left\{ W-\langle W\rangle ,P-\langle P\rangle \right\}  \label{zs}
\end{equation}
or by
\begin{equation}
F=i\left[ \left( 2a^{-}-\left\langle a^{-}\right\rangle \right) \left\langle
a^{-}\right\rangle +\left( -2a^{+}+\left\langle a^{+}\right\rangle \right)
\left\langle a^{+}\right\rangle -a^{-2}+a^{+2}\right]
\end{equation}
in terms of the operators $a^{-}$ and $a^{+}$.

The symbol $\left\{ ,\right\} $ in $\left( 12\right) $ stands for the
anti-commutator. When there is a correlation between $W$ and $P,$ i.e. $%
\langle F\rangle \neq 0,$ the relation $\left( 11\right) $ is a
generalization of the usual one (the Heisenberg uncertainty condition)
\begin{equation}
\left( \Delta W\right) ^2\left( \Delta P\right) ^2\geq \frac 14\langle
G\rangle ^2  \label{zd}
\end{equation}
The special from $\left( 14\right) $ is, of course, identical with the
general form $\left( 11\right) $ if $W$ and $P$ are uncorrelated, i.e. if $%
\langle F\rangle =0$. The general uncertainty relation $\left( 11\right) $
is better suited to determine the lower bound on the product of variances in
the measurement of observables corresponding to non-canonical operators. The
Robertson-Schr\"odinger uncertainty relation gives us a new understanding of
which states are coherent and which are squeezed for an arbitrary quantum
system. Indeed the so-called generalized intelligent states are obtained
when the equality in the Robertson-Schr\"odinger relation is realized $%
\left[ 26\right] $. The inequality in $\left( 11\right) $ becomes the
equality for the states $\left| \Psi \right\rangle $ satisfying the equation
\begin{equation}
\left( W+i\lambda P\right) \left| \Psi \right\rangle =z\sqrt{2}\left| \Psi
\right\rangle \hspace{1.0in}\lambda ,z\in \mathbf{C}  \label{zf}
\end{equation}
As a consequence, we have the following relations
\begin{equation}
\left( \Delta W\right) ^2=\left| \lambda \right| \Delta \hspace{1.0in}\left(
\Delta P\right) ^2=\frac 1{\left| \lambda \right| }\Delta  \label{zg}
\end{equation}
with
\begin{equation}
\Delta =\frac 12\sqrt{\left\langle G\right\rangle ^2+\left\langle
F\right\rangle ^2}  \label{zh}
\end{equation}
Note that the average values$\left\langle G\right\rangle $ and $\left\langle
F\right\rangle $, in the states satisfying the eigenvalue equation $\left(
15\right) ,$ can be expressed in terms of the variances as
\begin{eqnarray}
\left\langle G\right\rangle &=&2\hbox{Re}\left( \lambda \right) \left(
\Delta P\right) ^2  \nonumber \\
\left\langle F\right\rangle &=&2\hbox{Im}\left( \lambda \right) \left(
\Delta P\right) ^2  \label{zh}
\end{eqnarray}
It is clear, from $\left( 16\right) ,$ that if $\left| \lambda \right| =1$
we have

\begin{equation}
\left( \Delta W\right) ^2=\left( \Delta P\right) ^2  \label{zj}
\end{equation}
and we call the states satisfying $\left( 15\right) $, with $\left| \lambda
\right| =1,$ the generalized coherent states and; if $\left| \lambda \right|
\neq 1$, the states are called generalized squeezed states.

Using the equation $\left( 15\right) ,$ one can obtain some general
relations for the average values and dispersions for $W$ and $P$ in the
states which minimize the Robertson-Schr\"odinger uncertainty relation $%
\left( 11\right) $. Indeed, we have

\begin{eqnarray}
\left( \Delta W\right) ^2 &=&\frac 12\left( \hbox{Re}\left( \lambda \right)
\left\langle G\right\rangle +\hbox{Im}\left( \lambda \right) \left\langle
F\right\rangle \right)  \label{zk} \\
\left( \Delta P\right) ^2 &=&\frac 1{2\left| \lambda \right| ^2}\left( %
\hbox{Re}\left( \lambda \right) \left\langle G\right\rangle +\hbox{Im}\left(
\lambda \right) \left\langle F\right\rangle \right)  \label{zm} \\
\hbox{Im}\left( \lambda \right) \left\langle G\right\rangle &=&\hbox{Re}%
\left( \lambda \right) \left\langle F\right\rangle
\end{eqnarray}
We conclude this section by noticing that the minimization of the
Robertson-Schr\"odinger uncertainty relation leads to generalized
coherent states for $\left| \lambda \right| =1$(including the
so-called Gazeau-Klauder states obtained here for $\lambda =1$,
which minimize the Heisenberg uncertainty condition and are
eigenvectors of the annihilation operator $a^{-}$), and
generalized squeezed states for $\left| \lambda \right| \neq 1.$

\section{\textbf{Generalized Intelligent States}}

In the following, we will solve the eigenvalue equation $\left( 15\right) $
in order to give a complete classification of the coherent and squeezed
states for an arbitrary quantum system. To solve the eigenvalue equation $%
\left( 15\right) $, it is convenient to use the definition of $W$ and $P$ in
terms of the creation and annihilation operators $a^{+}$ and $a^{-}.$ The
equation $\left( 15\right) $ rewrite in the following form:
\begin{equation}
\left\{ \left( 1-\lambda \right) a^{+}+\left( 1+\lambda \right)
a^{-}\right\} \left| \Psi \right\rangle =2z\left| \Psi \right\rangle
\label{eq}
\end{equation}
Let us compute $\left| \Psi \right\rangle $ explicitly using $\left(
23\right) .$ We take

\begin{equation}
\left| \Psi \right\rangle =\sum\limits_{n=0}^\infty c_n\left| \psi
_n\right\rangle  \label{es}
\end{equation}
so that
\[
\left( 1-\lambda \right) c_{n-1}\sqrt{e_n}e^{-i\alpha \left(
e_n-e_{n-1}\right) }+\left( 1+\lambda \right) c_{n+1}\sqrt{e_{n+1}}%
e^{i\alpha \left( e_{n+1}-e_n\right) }=2zc_n
\]

\begin{equation}
\left( 1+\lambda \right) \sqrt{e_1}c_1=2ze^{-i\alpha e_1}c_0  \label{ed}
\end{equation}
Using the latter relation, let us give a complete classification of
generalized intelligent states for an arbitrary quantum system. We will
analyse the solution for the following cases: ($\lambda =1$, $z\neq 0$), ($%
\lambda =-1$, $z\neq 0$), ($\lambda \neq -1$, $z=0$) and ($\lambda \neq -1$,
$z\neq 0$). In each case, we give the solution of the equation $\left(
23\right) $ as some operator acting on the ground state $\left| \Psi
_0\right\rangle $ of the quantum system under consideration.

\subsection{Gazeau-Klauder Coherent states}

As we will see this set of states correspond to the situation where ($%
\lambda =1$, $z\neq 0$). In this case the equation $\left( 25\right) $
rewritten as
\[
c_n=z^n\frac{e^{-i\alpha e_n}}{\sqrt{e_ne_{n-1}\hbox{...}e_1}}c_0
\]
Then, the coherent states is given by

\begin{equation}
\left| \Psi \right\rangle =\left| z,\alpha \right\rangle
=c_0\sum\limits_{n=0}^\infty \frac{z^ne^{-i\alpha e_n}}{\sqrt{f\left(
n\right) }}\left| \psi _n\right\rangle  \label{eg}
\end{equation}
where the function $f\left( n\right) $ is defined by:
\begin{equation}
f\left( n\right) =\left\{
\begin{array}{c}
e_ne_{n-1}\hbox{...}e_1\hspace{1.0in}\hbox{for }n\neq 0 \\
1\hspace{1.0in}\hspace{1.6cm}\hbox{for }n=0
\end{array}
\right.  \label{eh}
\end{equation}
The normalization constant $c_0$ are calculated from the normalization
condition

\begin{equation}
\left\langle z,\alpha \right. \left| \hbox{ }z,\alpha \right\rangle =1
\label{ej}
\end{equation}
and is given by
\begin{equation}
c_0=\left( \sum\limits_{n=0}^\infty \frac{\left| z\right| ^{2n}}{f\left(
n\right) }\right) ^{-\frac 12}  \label{ej}
\end{equation}
the coherent states obtained here are, then, solutions of the eigenvalue
equation

\begin{equation}
a^{-}\left| \Psi \right\rangle =z\left| \Psi \right\rangle  \label{el}
\end{equation}
In other words, the states $\left| \Psi \right\rangle $ are the eigenstates
of the annihilation operator. This is one of the possibilities to define the
coherent states. It is well known that there are several non equivalent
definitions of them for a general system $\left[ 1,2\right] $. In the
arbitrary quantum system case, the connection with a possible
group-theoretical approach cannot be applied because, in contrast to the
harmonic oscillator, the operators $a^{+},$ $a^{-}$ and $G$ do not close a
Lie algebra. If we want to obtain the usual Heisenberg-Weyl algebra, we have
to modify these operators for new ones, which will be labelled $A^{-}$ and 
$%
A^{+}$ and satisfy the canonical commutation relation

\begin{equation}
\left[ A^{-},A^{+}\right] =1
\end{equation}
Let us take
\begin{equation}
A^{-}=a^{-}\hspace{1.0in}A^{+}=\frac N{g\left( N\right) }a^{+}
\end{equation}
where the operator $g\left( N\right) $ is defined by
\begin{equation}
g\left( N\right) =a^{+}a^{-}=H
\end{equation}
Note that $A^{-}$ and $A^{+}$ are not self-adjoint. So, it is possible to
rewrite $\left| \Psi \right\rangle $ (up to normalization) as
\begin{equation}
\left| \Psi \right\rangle =\exp \left( z\frac N{g\left( N\right)
}a^{+}\right) \left| \psi _0\right\rangle
\end{equation}
We see then that the coherent states ($\lambda =1$) minimize the Heisenberg
uncertainty relation, and are defined as eigenvectors of the annihilation
operator $a^{-}.$ They can also be given as the action of operator $\exp
\left( zA^{+}\right) $ on the ground state $\left| \psi _0\right\rangle 
$.\\%
It is easy to see that for $\lambda =1$, we have

\begin{equation}
\left( \Delta W\right) ^2=\left( \Delta P\right) ^2=\frac 12\left\langle
G\right\rangle  \label{rf}
\end{equation}
where
\begin{equation}
\left\langle G\right\rangle =c_0^2\sum\limits_{n=0}^\infty \frac{\left|
z\right| ^{2n}}{f\left( n\right) }e_{n+1}-\left| z\right| ^2
\end{equation}
and
\begin{equation}
\left\langle F\right\rangle =0  \label{rg}
\end{equation}
The latter equation traduces the fact that there is no correlation between 
$%
W $ and $P$.

Let us note that our coherent states coincide with the ones proposed by
Gazeau and Klauder $\left[ 14\right] ,$ where a set of four requirements of
such states has been imposed, i.e. continuity, resolution of unity, temporal
stability and action identity. Let us now verify that our coherent states
satisfy all these requirements. However, it should be noted that our
coherent states satisfy additional properties. They minimize the Heisenberg
uncertainty condition, are eigenstates of $a^{-}$ and are given as the
action of the operator $\exp \left( zA^{+}\right) $ on the ground state $%
\left| \psi _0\right\rangle $ (see equation. $\left( 34\right) $).

We see that they are continuous in $z\in \mathbf{C}$ and $\alpha \in 
\mathbf{%
R}$. Moreover, the presence of the phase factor in the definition $\left(
3\right) $ of the action of $a^{-}$ and $a^{+}$ leads to temporal stability
of the coherent states. Indeed, we have
\begin{equation}
e^{iHt}\left| z,\alpha \right\rangle =\left| z,\alpha +t\right\rangle
\end{equation}
The analysis of completeness (in fact, the overcompleteness) requires to
compute the resolution of the identity, that is
\begin{equation}
\int \left| z,\alpha \right\rangle \left\langle z,\alpha \right| \hbox{ }%
d\mu (z)=I_{\mathcal{H}}
\end{equation}
Note that the integral is over the disk \{$z\in \mathbf{C,}\left| z\right| 
<%
\mathcal{R}$\}, where the radius of convergence $\mathcal{R}$ is
\begin{equation}
\mathcal{R}\hbox{=}\lim\limits_{n\rightarrow \infty }\hbox{
}^n\sqrt{f(n)}
\end{equation}
and the measure $d\mu (z)$\ has to be determined. We suppose that $d\mu 
(z)$%
\ depends only on $\left| z\right| $.Then taking
\begin{equation}
d\mu (z)=\left[ c_0\right] ^{2}h(r^2)rdrd\phi \hspace{1.5cm}z=re^{i\phi }
\end{equation}
and using the coherent states (given by formula $\left( 26\right) $), we can
write $\left( 39\right) $ as
\begin{equation}
I_{\mathcal{H}}=\sum\limits_{n=0}^\infty \left| \psi _n\right\rangle
\left\langle \psi _n\right| \left[ \frac \pi {f(n)}\int_0^{\mathcal{R}%
^2}h(u)u^ndu\right]
\end{equation}
The resolution of the identity is then equivalent to the determination of
the function $h(u)$ satisfying
\begin{equation}
\int_0^{\mathcal{R}^2}h(u)\hbox{ }u^{n-1}du=\frac{f(n-1)}\pi
\end{equation}
For $\mathcal{R}$ $\rightarrow \infty $, it is clear that $h(u)$ is the
inverse Mellin transform of $\frac{f(n-1)}\pi $

\begin{equation}
h(u)=\frac 1{2\pi i}\int_{c-i\infty }^{c+i\infty }\frac{f(s-1)}\pi u^{-s}ds%
\hspace{1cm}c\in \mathbf{R}
\end{equation}
Note that explicit calculation of the function $h(u)$ requires the explicit
knowledge of the spectrum of the quantum system under consideration. The
measure of the coherent states is related to the spectrum and the special
application was treated for the P\"oschl--Teller potential $\left[ 17\right]
$, Morse potential $\left[ 29\right] $ and Jaynes-Cummings model $\left[
30\right] .$\\Using equation $\left( 30\right) $, one can obtain the mean
value of the Hamiltonian $H$ in the states $\left| z,\alpha \right\rangle $%
\begin{equation}
\left\langle z,\alpha \right| H\left| z,\alpha \right\rangle =\left|
z\right| ^2
\end{equation}
This relation is known as the action identity. It is clear now that our
coherent states satisfy also the Gazeau-Klauder requirements absolutely
necessary to define coherent states for an arbitrary quantum system.

\subsection{\textbf{The case (}$\mathbf{\lambda =-1,z\neq 0}$\textbf{)}}

In this case, we have to solve the eigenvalue equation

\begin{equation}
a^{+}\left| \Psi \right\rangle =z\left| \Psi \right\rangle  \label{qc}
\end{equation}
Then, the recurrence relation $\left( 25\right) $ rewritten

\begin{equation}
zc_n=c_{n-1}\sqrt{e_n}e^{-i\alpha \left( e_n-e_{n-1}\right) }\hspace{1.5cm}%
\hbox{and}\hspace{0.5cm}c_0=0  \label{qv}
\end{equation}
Then all coefficients vanish and we conclude that the solution in this case
cannot be normalized. The case $\lambda =-1$, leading to the unnormalized
solution, is not of interest.

\subsection{\textbf{The case (}$\mathbf{\lambda \neq -1,z=0}$)}

In the case where $\lambda \neq -1$ we will produce completely the set of
solutions which will give squeezed $\left( \left| \lambda \right| \neq
1\right) $ and generalized coherent states $\left( \left| \lambda \right|
=1\right) $. We start by examining the special case where $\lambda \neq -1$
and $z=0$ in order to have an idea about the general solution of the
eigenvalue equation in the general case corresponding to situation where $%
\lambda \neq -1$ and $z\neq 0$. Therefore, in the case $\lambda \neq -1$, $%
z=0,$ expanding the state $\left| \Psi \right\rangle $ as

\begin{equation}
\left| \Psi \right\rangle =\sum\limits_{n=0}^\infty c_n\left| \psi
_n\right\rangle  \label{qb}
\end{equation}
and using the equations $\left( 25\right) $, one can see that the
coefficients $c_n$ satisfy the following recurrence formulae

\begin{equation}
\left( 1+\lambda \right) c_{n+1}\sqrt{e_{n+1}}e^{i\alpha \left(
e_{n+1}-e_n\right) }=\left( \lambda -1\right) c_{n-1}\sqrt{e_n}e^{-i\alpha
\left( e_n-e_{n-1}\right) }  \label{qn}
\end{equation}
and

\begin{equation}
c_1=0  \label{qa}
\end{equation}
Then, the solution of the equation $\left( 23\right) $ is a linear
combination of the states $\left| \psi _{2k}\right\rangle $ $(k=0,1,2,3...)$

\begin{equation}
\left| \Psi \right\rangle =\sum\limits_{k=0}^\infty c_{2k}\left| \psi
_{2k}\right\rangle  \label{qz}
\end{equation}
where the coefficients $c_{2k}$ are given by

\begin{equation}
c_{2k}=\left( \frac{\lambda -1}{\lambda +1}\right) ^k\sqrt{\frac{%
e_1e_3...e_{2k-1}}{e_2e_4...e_{2k}}}e^{-i\alpha e_{2k}}c_0\hspace{1cm}k\geq 
1
\label{qe}
\end{equation}
(Note that the coefficients $c_{2k-1}=0$ for $k\geq 1.$) The coefficients $%
c_0$ can be calculated by imposing the normalization condition: %
\hspace{0.3cm}$\left\langle \Psi \right. \left| \Psi \right\rangle =1$. We
obtain

\begin{equation}
c_0=\left[ \sum\limits_{k=1}^\infty \left| \frac{\lambda -1}{\lambda +1}%
\right| ^{2k}\frac{\left( e_1e_3...e_{2k-1}\right) ^2}{f(2k)}\right]
^{-\frac 12}\hspace{1cm}k\geq 1  \label{qr}
\end{equation}
It is interesting to mention that the state $\left| \Psi \right\rangle $ can
be obtained by acting of the operator

\begin{equation}
U\left( \lambda \neq -1,z=0\right) =c_0\exp \left( \frac 12\left( \frac{%
\lambda -1}{\lambda +1}\right) \frac N{g\left( N\right) }\left( a^{+}\right)
^2\right)  \label{qt}
\end{equation}
on the ground state $\left| \psi _0\right\rangle $ where $g\left( N\right) $
is defined as in $\left( 33\right) $. So, the states minimizing the
Robertson-Schr\"odinger uncertainty relation with $\lambda \neq -1$ and 
$z=0$%
, are given by:

\begin{equation}
\left| \Psi \right\rangle =U\left( \lambda \neq -1,z=0\right) \left| \psi
_0\right\rangle  \label{qy}
\end{equation}
Note that for $\lambda =1$, we have
\begin{equation}
U(\lambda =1\hbox{, }z=0)=c_0
\end{equation}
and the states $\left| \Psi \right\rangle $ are nothing but the ground state
$\left| \psi _0\right\rangle $ which is annihilated by the operator $a^{-}$ 
(%
$a^{-}\left| \psi _0\right\rangle =0$)

The result of this subsection can be seen as a first step to obtain the
generalized intelligent states for an arbitrary quantum system.

\subsection{\textbf{The case (}$\mathbf{\lambda \neq -1},$\textbf{\ }$%
\mathbf{z\neq 0}$)}

This case is more interesting and leads to squeezed states for an arbitrary
quantum system. We start by solving the eigenvalue equation $\left(
23\right) $ and we give the solution of this equation as the action of some
operator, which will be defined later on, on the ground state of the system
under consideration. The example of the harmonic oscillator is discussed in
the end of this section.

In the case where $\lambda \neq -1$ and\textbf{\ }$z\neq 0$, the
eigenvalue equation $\left( 23\right) $ gives the following
recurrence formulae

\begin{equation}
\left( 1-\lambda \right) c_{n-1}\sqrt{e_n}e^{-i\alpha \left(
e_n-e_{n-1}\right) }+\left( 1+\lambda \right) c_{n+1}\sqrt{e_{n+1}}%
e^{i\alpha \left( e_{n+1}-e_n\right) }=2zc_n  \label{qi}
\end{equation}
and

\begin{equation}
c_1=\frac{2ze^{-i\alpha e_1}}{\left( 1+\lambda \right) \sqrt{e_1}}c_0
\label{qi}
\end{equation}
Setting:

\begin{equation}
A_{n+1}=\frac{c_{n+1}}{c_n}\sqrt{e_{n+1}}e^{i\alpha \left(
e_{n+1}-e_n\right) }  \label{qo}
\end{equation}
The relations $\left( 58\right) $ and $\left( 59\right) $ can be written in
the following form:

\begin{equation}
A_1=\frac{2z}{\left( 1+\lambda \right) }\hspace{0.5cm}\hbox{ and }%
\hspace{1cm}A_n=\frac{2z}{\left( 1+\lambda \right) }+\left( \frac{\lambda 
-1%
}{\lambda +1}\right) \frac{e_{n-1}}{A_{n-1}}  \label{qp}
\end{equation}
>From the latter equations, we obtain the coefficients $A_n$ which are
expressed as continued fraction. They are given by

\begin{equation}
A_n=\frac{2z}{1+\lambda }+\frac{\left( \frac{\lambda -1}{\lambda +1}\right)
e_{n-1}}{\frac{2z}{1+\lambda }+\frac{\left( \frac{\lambda -1}{\lambda +1}%
\right) e_{n-2}}{\frac{2z}{1+\lambda }+\frac{\left( \frac{\lambda -1}{%
\lambda +1}\right) e_{n-3}}{%
\begin{array}{c}
\frac{2z}{1+\lambda }+.....\hspace{4cm} \\
\hspace{0.4cm}.+..... \\
\hspace{5cm}\frac{2z}{1+\lambda }+\frac{\left( \frac{\lambda -1}{\lambda 
+1}%
\right) e_1}{A_1}
\end{array}
}}}  \label{wq}
\end{equation}
Now we are able to compute the coefficients $c_n$. Indeed, they are given by
the following expression

\begin{equation}
c_n=c_0\frac{\left( 2z\right) ^n}{\left( 1+\lambda \right) ^n\sqrt{f\left(
n\right) }}\left[ \sum\limits_{h=0\left( 1\right) \left[ \frac n2\right]
}\left( -1\right) ^h\frac{\left( 1-\lambda ^2\right) ^h}{\left( 2z\right)
^{2h}}\Delta \left( n,h\right) \right] e^{-i\alpha e_n}  \label{cq}
\end{equation}
where the symbol $\left[ \frac n2\right] $ represents the integer part of $%
\frac n2$ and the function $\Delta \left( n,h\right) $ is defined by

\begin{equation}
\Delta \left( n,h\right) =\sum\limits_{j_1=1}^{n-\left( 2h-1\right)
}e_{j_1}\left[ \sum\limits_{j_2=j_1+2}^{n-\left( 2h-3\right)
}e_{j_2}...\left[ ...\left[ \sum\limits_{j_h=j_{h-1}+2}^{n-1}e_{j_h}\right]
\right] ...\right]  \label{vq}
\end{equation}
As an example of computation of the $c_n$, we give the first four
coefficients

\begin{eqnarray*}
c_1 &=&\frac{2z}{\left( 1+\lambda \right) \sqrt{e_1}}e^{-i\alpha e_1}c_0 \\
c_2 &=&\frac{\left( 2z\right) ^2}{\left( 1+\lambda \right) ^2\sqrt{e_1e_2}}%
\left[ 1+\frac{\lambda ^2-1}{\left( 2z\right) ^2}e_1\right] e^{-i\alpha
e_2}c_0 \\
c_3 &=&\frac{\left( 2z\right) ^3}{\left( 1+\lambda \right) 
^3\sqrt{e_1e_2e_3}%
}\left[ 1+\frac{\lambda ^2-1}{\left( 2z\right) ^2}\left( e_1+e_2\right)
\right] e^{-i\alpha e_3}c_0 \\
c_4 &=&\frac{\left( 2z\right) ^4}{\left( 1+\lambda \right) ^4\sqrt{%
e_1e_2e_3e_4}}\left[ 1+\frac{\lambda ^2-1}{\left( 2z\right) ^2}\left(
e_1+e_2+e_3\right) +\left( \frac{\lambda ^2-1}{\left( 2z\right) ^2}\right)
^2e_1e_3\right] e^{-i\alpha e_4}c_0
\end{eqnarray*}
As we mentioned in the beginning of this section, the general solution of
the eigenvalue equation $\left( 23\right) $ can be written as the action of
some operator on the ground state $\left| \psi _0\right\rangle $ of $H.$ A
more or less complicated calculation give the following result

\begin{equation}
\left| \Psi \right\rangle =U\left( \lambda \neq -1,z\neq 0\right) \left|
\psi _0\right\rangle
\end{equation}
where

\begin{equation}
U\left( \lambda \neq -1,z\neq 0\right) =c_0\sum\limits_{n=0}^\infty \left(
\left( \frac{2z}{\lambda +1}\right) \frac{a^{+}}{g\left( N\right) }+\left(
\frac{\lambda -1}{\lambda +1}\right) \frac 1{g\left( N\right) }\left(
a^{+}\right) ^2\right) ^n
\end{equation}
In the case where $\lambda =1,$ we recover the operator which acting on $%
\left| \psi _0\right\rangle ,$ gives the Gazeau-Klauder coherent states.

Note also that for $\lambda \neq -1$ and $z=0$, the operator $\left(
65\right) $ coincides with that given by $\left( 54\right) $. Moreover it is
not difficult to see that the generalized intelligent states are stable
temporally. Finally, as a first illustration of our construction, we can
obtain the generalized intelligent states for the standard harmonic
oscillator. Indeed, using the equations $\left( 64\right) $ and $\left(
65\right) $ and setting $g(N)=N$, we have
\begin{equation}
\left| \Psi \right\rangle =c_0\exp \left[ \left( \frac{\lambda -1}{\lambda 
+1%
}\right) \frac{\left( a^{+}\right) ^2}2\right] \exp \left[ \left( 
\frac{2z}{%
\lambda +1}\right) a^{+}\right] \left| 0\right\rangle
\end{equation}
where $\left| 0\right\rangle $ is the ground state for the harmonic
oscillator.

\section{Fock-Bargmann representation}

It is well known that the Fock-Bargmann representation enables one to find
simpler solutions of a number of problems, exploiting the theory of
analytical entire functions.\\In this part of our work, generalizing the
pioneering work of Bargmann $\left[ 31\right] $ for the usual harmonic
oscillator, we will study the Fock-Bargmann representation of the dynamical
algebra generated by annihilation and creation operators corresponding to an
arbitrary quantum system. We recall that in the Fock-Bargmann representation
for the harmonic oscillator, the creation operator $a^{+}$ is the
multiplication by $z$, while the operator $a^{-}$ is the differentiation
with the respect to $z.$

We define the Fock-Bargmann space as a space of functions $S$ which are
holomorphic\textrm{\ }on a ring $D$ of the complex plane. The scalar product
is written with an integral of the form

\begin{equation}
\left\langle f_1\right. \left| f_2\right\rangle =\int \overline{f_1\left(
z\right) }f_2\left( z\right) \hbox{ }d\mu \left( z\right)  \label{sd}
\end{equation}
Where $d\mu \left( z\right) $ is the measure defined above (see equation $%
\left( 41\right) $). The Fock-Bargmann representation of the dynamical
algebra $\left\{ a^{+},a^{-},G\right\} $is a representation on Fock-Bargmann
space such that the annihilation and the creation operators admit
eigenvectors generating $S$.\\Let $\left| h\right\rangle $ be a state of the
Hilbert space $\mathcal{H}$

\begin{equation}
\left| h\right\rangle =\sum\limits_{n=0}^\infty h_n\left| \psi
_n\right\rangle \hspace{1.0in}\sum\limits_{n=0}^\infty \left| h_n\right|
^2<\infty  \label{rt}
\end{equation}
Following the construction of $\left[ 31\right] $, any state $\left|
h\right\rangle $ of $\mathcal{H}$ in the Fock-Bargmann representation is
represented by a function of the complex variable $z$ (using the coherent
states associated with an arbitrary quantum system)

\begin{equation}
h\left( z\right) = \left\langle \overline{z},\alpha \right. \left|
h\right\rangle =\sum\limits_{n=0}^\infty \frac{z^ne^{i\alpha e_n}}{\sqrt{%
f\left( n\right) }}h_n  \label{er}
\end{equation}
where the variable $z$ belongs to the domain $D$ of definition of the
eigenvalues of $a^{-}$ (annihilation operator). In particular, to the basis
vectors $\left| \psi _n\right\rangle $ there correspond the monominals

\begin{equation}
\psi _n(z)=\left\langle \overline{z},\alpha \right. \left| \psi
_n\right\rangle =\frac{z^ne^{i\alpha e_n}}{\sqrt{f\left( n\right)
}} \label{fg}
\end{equation}
Using the equations $\left( 69\right) $ and $\left( 70\right) $, we can
prove easily the following result.\\In the Fock-Bargmann representation, we
realize the annihilation operator $a^{-}$ by
\begin{equation}
a^{-}=z^{-1}g\left( z\frac d{dz}\right) \hspace{0.5cm},
\end{equation}
the creation operator $a^{+}$by
\begin{equation}
a^{+}=z\hspace{0.5cm},
\end{equation}
and the operator number by
\begin{equation}
N=z\frac d{dz}
\end{equation}
The Fock-Bargmann representation exists if we have a measure such that

\begin{equation}
\int \left| z,\alpha \right\rangle \left\langle z,\alpha \right| \hbox{ }%
d\mu \left( z\right) =I_{\mathcal{H}}  \label{ui}
\end{equation}
The existence of the measure, which was discussed previously for an
arbitrary quantum system, ensures that the scalar product takes the form $%
\left( 67\right) $\\We note that in the case where
\begin{equation}
g\left( z\frac d{dz}\right) =z\frac d{dz}\hspace{1cm},\hspace{0.6cm}%
\hbox{i.e}\hspace{1cm}g\left( N\right) =N  \label{ds}
\end{equation}
we recover the well-known Fock-Bargmann representation of the harmonic
oscillator.\\An interesting case concerns the situation where
\begin{equation}
g\left( N\right) =N\left( N+\upsilon \right)  \label{vb}
\end{equation}
which occurs, for instance, when one deals with a quantum system
evolving in the infinite square well and P\"oschl-Teller
potentials. The Fock-Bargmann realization presented in this
section will be the corner stone to construct the generalized
intelligent states for these potentials. This matter will be
considered in section 5.

\section{P\"oschl-Teller Intelligent States}

We start by recalling the eigenvalues and eigenvectors of infinite square
well and P\"oschl-Teller potentials (cf section 5.1). We discuss the
Gazeau-Klauder coherent states associated with these two quantum systems (cf
section 5.2) and, using the Fock-Bargmann representation, we give an
analytic realization of the generalized intelligent states corresponding to
infinite square well and P\"oschl-Teller potentials (cf section 5.3).

\subsection{Spectrum of the P\"oschl-Teller potentials}

We consider the Hamiltonian

\begin{equation}
H=-\frac{d^2}{dx^2}+V_{\kappa ,\lambda }\left( x\right)  \label{xs}
\end{equation}
describing a parlicle, on the line, and subjected to the potential

\begin{equation}
V_{\kappa ,\lambda }\left( x\right) =\left\{
\begin{array}{c}
\frac 1{4a^2}\left[ \frac{\kappa \left( \kappa -1\right) }{\sin ^2\left(
\frac x{2a}\right) }+\frac{\lambda \left( \lambda -1\right) }{\cos ^2\left(
\frac x{2a}\right) }\right] -\frac{\left( \lambda +\kappa \right) ^2}{4a^2}%
\hspace{1cm}0<x<\pi a \\
\hspace{1cm}\infty \hspace{2.5cm}\hspace{2.5cm}x\leq 0\hspace{0.5cm},%
\hspace{0.5cm}x\geq \pi a
\end{array}
\right.  \label{dsf}
\end{equation}
for $\lambda >1$ and $\kappa >1$ (the parameter $\lambda $ should not be
confused with one appearing in the eigenvalue equation $\left( 15\right) 
$)$%
. $ It is well known that the P\"oschl-Teller potential $\left[ 32\right] $
interpolates between the harmonic oscillator and the infinite square well.
The infinite well takes place at the limit $\lambda =\kappa =1.$ The
Hamiltonian $H$ can be written in the following form

\begin{equation}
H=a_{\kappa ,\lambda }^{+}a_{\kappa ,\lambda }^{-}  \label{ssf}
\end{equation}
where the annihilation and creation operators are given by

\begin{equation}
a_{\kappa ,\lambda }^{\pm }=\left( \mp \frac d{dx}+W_{\kappa ,\lambda
}\left( x\right) \right)  \label{hj}
\end{equation}
in terms of the superpotentials $W_{\kappa ,\lambda }\left( x\right) $

\begin{equation}
W_{\kappa ,\lambda }\left( x\right) =\frac 1{2a}\left[ \kappa \hbox{cotg}%
\left( \frac x{2a}\right) -\lambda \tan \left( \frac x{2a}\right) \right]
\label{tu}
\end{equation}
The eigenvectors are given by

\begin{equation}
\psi _n\left( x\right) =\left[ c_n\left( \kappa ,\lambda \right) \right]
^{-\frac 12}\left( \cos \frac x{2a}\right) ^\lambda \left( \sin \frac
x{2a}\right) ^\kappa \frac{n!\Gamma \left( \kappa +\frac 12\right) }{\Gamma
\left( n+\kappa +\frac 12\right) }P_n^{\left( \kappa -\frac 12,\lambda
-\frac 12\right) }\left( \cos \frac xa\right)  \label{ty}
\end{equation}
where the normalization constant above is

\begin{equation}
c_n\left( \kappa ,\lambda \right) =a\frac{\Gamma \left( n+1\right) \hbox{ }%
\Gamma \left( \kappa +\frac 12\right) ^2\Gamma \left( n+\lambda +\frac
12\right) }{\Gamma \left( n+\kappa +\frac 12\right) \Gamma \left( n+\kappa
+\lambda \right) \Gamma \left( 2n+\kappa +\lambda \right) }  \label{hhj}
\end{equation}
The eigenvalues of $H$ are given by

\begin{equation}
H\left| \psi _n\right\rangle =n\left( n+\kappa +\lambda \right) \left| \psi
_n\right\rangle  \label{gfjk}
\end{equation}
The creation and annihilation operators $a_{k,\lambda }^{+}$ and $%
a_{k,\lambda }^{-}$ act on $\left| \psi _n\right\rangle $ as follows

\begin{eqnarray}
a_{\kappa ,\lambda }^{+}\left| \psi _n\right\rangle &=&\sqrt{\left(
n+1\right) \left( n+1+\kappa +\lambda \right) }e^{-i\alpha \left(
2n+1+\kappa +\lambda \right) }\left| \psi _{n+1}\right\rangle  \label{nb} \\
a_{\kappa ,\lambda }^{-}\left| \psi _n\right\rangle &=&\sqrt{n\left(
n+\kappa +\lambda \right) }e^{i\alpha \left( 2n-1+\kappa +\lambda \right)
}\left| \psi _{n-1}\right\rangle  \nonumber
\end{eqnarray}
>From the latter equation, one can verify that $a_{k,\lambda }^{+}$ and $%
a_{k,\lambda }^{-}$ satisfy the following commutations relations

\begin{equation}
\left[ a_{\kappa ,\lambda }^{-},a_{\kappa ,\lambda }^{+}\right] =G_{\kappa
,\lambda }\left( N\right)  \label{juk}
\end{equation}
where
\begin{equation}
G_{\kappa ,\lambda }\left( N\right) \equiv G=2N+\left( 1+\kappa +\lambda
\right)  \label{tyu}
\end{equation}
and the operator $N$ is defined, as in the first section, by
\begin{equation}
N\left| \psi _n\right\rangle =n\left| \psi _n\right\rangle  \label{bnv}
\end{equation}
Here also, we mention that $N\neq a_{\kappa ,\lambda }^{+}a_{\kappa ,\lambda
}^{-}=H$.

\subsection{Coherent states for the P\"oschl-Teller potentials}

>From the result obtained in section 3, the coherent states read as
\begin{equation}
\left| z,\alpha \right\rangle =\mathcal{N}\left( \left| z\right| \right)
\sum\limits_{n=0}^\infty \frac{z^ne^{-i\alpha n(n+\kappa +\lambda 
)}}{\sqrt{%
\Gamma (n+1)\Gamma (n+\kappa +\lambda +1)}}\left| \psi _n\right\rangle
\label{tuyg}
\end{equation}
The normalization constant is given by:
\begin{equation}
\mathcal{N}\left( \left| z\right| \right) =\sqrt{\frac{\left| z\right|
^{\kappa +\lambda }}{I_{\kappa +\lambda }\left( 2\left| z\right| \right) }}
\label{RFD}
\end{equation}
It is easy to verify that the radius of convergence $\mathcal{R}$ is
infinite.\\The identity resolution is explicitly given by
\begin{equation}
\int \left| z,\alpha \right\rangle \left\langle z,\alpha \right| d\mu \left(
z\right) =I_{\mathcal{H}}  \label{dgh}
\end{equation}
where the measure $d\mu \left( z\right) $ can be computed by the inverse
Mellin transform $\left[ 33\right] $%
\begin{equation}
d\mu \left( z\right) =\frac 2\pi I_{\kappa +\lambda }\left( 2r\right) K_{%
\frac{\kappa +\lambda }2}\left( 2r\right) rdrd\phi \hspace{1.5cm}z=re^{i\phi
}  \label{gjh}
\end{equation}
The coherent states of the infinite square well are obtained from the
P\"oschl-Teller ones simply by putting $\lambda +\kappa =2$

The coherent states from an overcomplete family of states (resolving the
identity resolution by integration with respect to measure given by $\left(
92\right) $), and provide a representation of any state $\left| \Psi
\right\rangle $ by an entire analytic function $\left\langle \Psi \right.
\left| z,\alpha \right\rangle .$\\Using the Fock-Bargmann representation
discussed above, the creation and annihilation operators, corresponding to
the quantum system evolving in P\"oschl-Teller (or in the infinite square
well) potentials, are realized by

\begin{equation}
a_{\kappa ,\lambda }^{+}=z\hspace{1cm}\hbox{,}\hspace{2cm}\hbox{
}a_{\kappa ,\lambda }^{-}=z\frac{d^2}{dz^2}+\left( \lambda +\kappa +1\right)
\frac d{dz}  \label{fgh}
\end{equation}
It is easy to see that the operator $G$, in this representation, act as
\begin{equation}
G=2z\frac d{dz}+\left( \lambda +\kappa +1\right) \hspace{1cm}  \label{gh}
\end{equation}
We will use this representation to construct P\"oschl-Teller generalized
intelligent states. The ones corresponding to infinite square well can be
obtained simply by taking $\lambda =\kappa =1$.

\subsection{P\"oschl-Teller Generalized Intelligent States}

We consider the P\"oschl-Teller equal variance $\left| z,\alpha
\right\rangle $ (the eigenstates of the annihilation operator $a_{\kappa
,\lambda }^{-}$). As we mentioned before, these states provide a
representation of any state $\left| \Phi \right\rangle $ (belonging to the
Hilbert space corresponding to the system evolving in the P\"oschl-Teller
potential) by an entire analytic function $\left\langle \Phi \right. \left|
z,\alpha \right\rangle $. Then, by means of the analytical realization, and
using the differential representation of the creation and annihilation
operators, we can construct the P\"oschl-Teller generalized intelligent
states $\left| z^{\prime },\lambda ,\alpha \right\rangle $ (we denote for a
while the eigenvalue by $z^{\prime }$ and we put $\lambda +\kappa =\upsilon 
$%
).\\The eigenvalue equation $\left( 23\right) $ which takes the form
\begin{equation}
\left[ \left( 1+\lambda \right) a^{-}+\left( 1-\lambda \right) a^{+}\right]
\left| z^{\prime },\lambda ,\alpha \right\rangle =2z^{\prime }\left|
z^{\prime },\lambda ,\alpha \right\rangle  \label{gn}
\end{equation}
now reads
\begin{equation}
\left[ \left( 1+\lambda \right) \left( z\frac{d^2}{dz^2}+\left( \upsilon
+1\right) \frac d{dz}\right) +\left( 1-\lambda \right) z\right] \Phi
_{\left( z^{\prime },\lambda ,\alpha \right) }\left( z\right) =2z^{\prime
}\Phi _{\left( z^{\prime },\lambda ,\alpha \right) }\left( z\right)
\label{gjf}
\end{equation}
By means of simple substitution, the above equation is reduced to the Kummer
equation for the confluent hypergeometric function $_1F_1\left( a,b,z\right)
$, so that we have the following solution of the equation $\left( 96\right) 
$
\begin{equation}
\Phi _{\left( z^{\prime },\lambda ,\alpha \right) }\left( z\right) =\exp
\left( \sqrt{\frac{\lambda -1}{\lambda +1}}z\right) \hbox{
}_1F_1\left( a,b,-2\sqrt{\frac{\lambda -1}{\lambda +1}}z\right)  \label{fsf}
\end{equation}
where
\begin{equation}
a=\frac{\upsilon +1}2-\frac{z^{\prime }}{\sqrt{\left( \lambda ^2-1\right) 
}}%
\hbox{\hspace{1.5cm}and }\hspace{1cm}b=\upsilon +1  \label{fg}
\end{equation}
Using the properties of the above hypergeometric function (cf equation $%
\left( 97\right) $), we arrive at the conclusion that the
squeezing parameter $\lambda $ obeys to the following condition
\begin{equation}
\sqrt{\left| \frac{1-\lambda }{1+\lambda }\right| }<1\hspace{1cm}%
\Longleftrightarrow \hspace{1cm}Re\lambda >0  \label{kkl}
\end{equation}
which is exactly the restriction on $\lambda $ imposed by the positivity of
the commutator $\left[ a^{-},a^{+}\right] =G\left( N\right) $ (see equation 
$%
\left( 18\right) $). Thus we obtain the P\"oschl-Teller generalized
Intelligent states in the coherent states representation in the form (up to
the normalization constant)

\begin{equation}
\left\langle z^{\prime },\lambda ,\alpha \right| \left. z,\alpha
\right\rangle =\exp \left( c^{*}z\right) \hbox{ }_1F_1\left(
a^{*},b,-2c^{*}z\right)  \label{rg}
\end{equation}
where
\begin{equation}
c=\sqrt{\frac{\lambda -1}{\lambda +1}}  \label{sd}
\end{equation}
and the parameters $a$ and $b$ are defined in formulae $\left( 98\right) $.
In the case where $\lambda =1$ (i.e. $c=0$), using the power series of $%
_1F_1\left( \alpha ,\beta ,x\right) $%
\begin{equation}
_1F_1\left( \alpha ,\beta ,x\right) =\sum\limits_{n=0}^\infty \frac{\left(
\alpha \right) _n}{\left( \beta \right) _n}\frac{x^n}{n!}\hspace{1cm}%
\hbox{where }\left( \alpha \right) _n=\alpha \left( \alpha +1\right)
...\left( \alpha +n-1\right)  \label{sdf}
\end{equation}
we obtain
\begin{equation}
\left\langle z,\lambda =1,\alpha \right| \left. z\right\rangle =\hbox{ }%
_0F_1\left( \upsilon +1,z\overline{z^{\prime }}\right)  \label{dfg}
\end{equation}
where
\begin{equation}
_0F_1\left( \alpha ,x\right) =\sum\limits_{n=0}^\infty \frac 1{\left( \alpha
\right) _n}\frac{x^n}{n!}  \label{dh}
\end{equation}
The result $\left( 103\right) $ coincides with the solution $\left(
89\right) $ for $\lambda =1,$ and we recover the P\"oschl-Teller coherent
states defined as the $a_{\kappa ,\lambda }^{-}$ eigenvectors.\\To close
this section, we discuss the coherence and squeezing of P\"oschl-Teller
generalized intelligent states.

As discussed above, in the case where $\lambda =1$, we have the so-called
Gazeau-Klauder coherent states. The dispersions $\Delta W\equiv \Delta
W_{\kappa ,\lambda }$ and $\Delta P$ saturate the Robertson-Schr\"odinger
uncertainty relation and we obtain
\begin{equation}
\left( \Delta W\right) ^2=\left( \Delta P\right) ^2=\frac 12\left\langle
G\right\rangle
\end{equation}
and $\left\langle F\right\rangle =0$. Using the expression of the operator 
$%
G $ in the case of P\"oschl-Teller (or infinite square well) potentials, one
can show that its mean value on the coherent states $\left| z,\alpha
\right\rangle $ is given by:
\begin{equation}
\left\langle G\right\rangle =\left\langle z,\alpha \right| G(N)\left|
z,\alpha \right\rangle =\left( 1+\upsilon \right) +\frac{2\left| z\right| 
^2%
}{\left( 1+\upsilon \right) }\frac{_0F_1\left( 2+\upsilon ,\left| z\right|
^2\right) }{_0F_1\left( 1+\upsilon ,\left| z\right| ^2\right) }
\end{equation}
Note also that
\begin{equation}
\left\langle G\right\rangle \geq 1+\upsilon \geq \frac 12
\end{equation}
which traduce the fact that the dispersions $\Delta P$ and $\Delta
W$ are greater than $\frac 1{{2}}$ (remember that in the case of
the harmonic oscillator we have $\Delta P=\Delta W=\frac 1{{2}}$).
This result constitutes a main difference with the well known
harmonic oscillator coherent states.

Another interesting situation concerns the case $\left| \lambda \right| =1$
with $\lambda \neq \pm 1$. The case $\lambda =1$ was discussed before and $%
\lambda =-1$ is not allowed by our construction. Taking $\lambda =e^{i\theta
}$ ($\theta \neq k\pi $ ; $k\in \mathbf{N}$). The states are coherents and
dispersions are given by

\begin{equation}
\left( \Delta W\right) ^2=\left( \Delta P\right) ^2=\frac 1{2\left| \cos
\theta \right| }\left\langle G\right\rangle
\end{equation}
The mean value of the operator $F$ is non-vanishing (vanishing only in the
Gazeau-Klauder coherent states). It is given by

\begin{equation}
\left\langle F\right\rangle =\hbox{tg}\theta \left\langle G\right\rangle
\end{equation}
>From the latter equation, we conclude that the presence of the correlation
do not forbid the system to be prepared in a coherent state. This result is
true for any quantum system. The properties of generalized intelligent
states turn out to be sensitive to the spectral properties of the commutator
$\left[ a_{\kappa ,\lambda }^{-},a_{\kappa ,\lambda }^{+}\right] =G(N)$.

Consider now (for the sake of completeness) the exceptional case of states
which minimize the Robertson-Schr\"odinger uncertainty relation with Re$%
\lambda =0$. In this case, we have eigenstates with vanishing mean value of 
$%
G$. In the same way, the mean value of $F$, on the generalized intelligent
states with Im$\lambda =0$, is zero. Finally, in the case where $\left|
\lambda \right| \neq 1,$ the generalized intelligent states exhibit strong
squeezing. This takes place, for example, when $\lambda \rightarrow 0$ (cf
equation $\left( 16\right) $) which can be easily derived even without
explicit calculation of the variances.

\section{Conclusion}

In this paper, we gave a complete classification of the eigenstates of the
eigenvalue equation arising from the minimization of the
Robertson-Schr\"odinger uncertainty relation. We obtained the so-called
generalized squeezed states for $\left| \lambda \right| \neq 1$ and the
generalized coherent states for $\left| \lambda \right| =1$. The latter
class includes the Gazeau-Klauder coherent states for which we examined the
properties known for them such as continuity, temporal stability, action
identity, and resolution of unity. The measure, which ensures the
overcompleteness of coherent states, is strongly related to the nature of
the spectrum under study. We also purposed, a Fock-Bargmann representation
of the creation and annihilation operators for an arbitrary quantum
system.This representation was useful to construct, in an analytical way the
generalized intelligent states for the P\"oschl-Teller and infinite square
well potentials. The results obtained through this work constitute a first
step to get more information about the squeezing and coherence for an
arbitrary quantum system and we believe that there is many directions on
this subject which can be explored. Indeed, we think that our results can be
adapted to the $x^4$-anharmonic oscillator $\left[ 34\right] .$ We also hope
to construct the Perelomov coherent states type (group theoretical approach)
for an arbitrary quantum system and compare them with Gazeau-Klauder ones.
These matters will be considered in a forthcoming work.

\newpage\

\end{document}